\documentstyle[preprint,prc,aps]{revtex}
\begin{document}
\title{HEAVY HEXAQUARKS IN A CHIRAL CONSTITUENT QUARK MODEL}
\author{S. Pepin\footnote{present address : Theoretical Physics Group,
Department of Physics and Astronomy, University of Manchester, Manchester
 M13 9PL, United Kingdom} and Fl. Stancu}
\address{Universit\'{e} de Li\`ege, Institut de Physique B.5, Sart Tilman,
B-4000 Li\`ege 1, Belgium}
\date{\today}
\maketitle
\begin{abstract}
We discuss the stability of hexaquark systems of type uuddsQ (Q=c or b)
within a chiral constituent quark model which successfully describes the baryon
spectra including the charmed ones. We find these systems highly unstable 
against strong decays and
give a comparison with some of the previous literature.
\end{abstract}

\section{INTRODUCTION}
According to QCD rules, systems such as tetraquarks, pentaquarks or hexaquarks
can, in principle, exist.
Their study is important in disentangling
between various QCD inspired models. Here we are mainly concerned with 
non-relativistic models, which simulate the low-energy limit of QCD.
In these models, the central part of the interquark potential usually 
contains a linear term which describes the QCD confinement and a Coulomb term
generated by the long-range one-gluon exchange (OGE) interaction. The
spin part is usually described by the chromo-magnetic part of the one-gluon 
exchange interaction, analogous to the Fermi-Breit interaction of QED  
\cite{RGG75,Ba88}.

An interest in the constituent quark model has recently been revitalized 
\cite{GR96} after recognition of the role of the spontaneous chiral symmetry 
of the QCD vacuum. This implies that the valence quarks acquire a constituent 
dynamical mass related to the quark condensate $<q\bar{q}>$ and that the
Goldstone bosons $\pi, K, \eta$ couple directly to constituents quarks 
\cite{We79}. It has been shown that the hyperfine splitting and especially 
the correct ordering of positive and negative parity states of baryons with
u,d and s quarks are produced by the short-range part of the Goldstone boson 
exchange (GBE) interaction \cite{GR96,G97,semrel}, instead of the OGE
interaction.

In Ref.\cite{SPG97}, we studied the stability of the H-particle, a uuddss
system, with $J^P=0^+$ and I=0, in the
frame of the chiral constituent quark model of Ref.\cite{GPP96}. We found
that the GBE interaction induces a strong repulsion in the flavour singlet
uuddss system with $J^P=0^+$ and I=0, i.e. this system lies 847 MeV above the
$\Lambda\Lambda$ threshold. This implies that the model of Ref.\cite{GPP96}
predicts that the H-particle should not exist, in contrast to Jaffe's 
\cite{Ja77} or many other studies based on conventional one-gluon exchange 
models \cite{allH}. In the
model used by Jaffe, the chromomagnetic interaction, gave more attraction
for the flavour singlet state than for two well separated
lambda baryons. In Jaffe's picture the H-particle should be a
compact object, in contrast to the molecular-type deuteron.

  In a recent study by Lichtenberg, Roncaglia and Predazzi \cite{LRP96},
the uuddss system is discussed in the context of a diquark model
and it is found unstable, in contrast to Jaffe's result. The above
authors also discuss hexaquarks in which one of the s quarks
is replaced by a c quark, the $H_c$ particle, or by a b quark, the $H_b$
particle. The charmed hexaquark is found unstable but the bottom
hexaquark is found stable by about 10 MeV with respect to the $\Lambda +
\Lambda_b$ threshold. Based on the concept of dynamical hadron 
supersymmetry, Ref.
\cite{LRP96} predicts the exotic masses from the ones of ordinary mesons and 
baryons as input, with no free parameters. On the other hand, more 
sophisticated calculations within a constituent quark model with 
chromomagnetic interaction give both $H_c$(I=0, J=3) and $H_b$(I=0, J=2 or 3)
stable by 7.7 MeV up to 13.8 MeV \cite{LSB93}. In fact,
from general grounds \cite{JMR94,MW93} one expects that the
stability of multiquark systems should increase with the mass asymetry of the
constituent quarks and this was tested for tetraquarks systems, in a 
conventional constituent quark model with chromomagnetic interaction
\cite{ZSGR86,SBS93,BS97}.

   Here we focus our attention
on hexaquarks and in particular on the uuddsQ (Q=c or b) system, which results
as a promising candidate from the above models.  We study 
the stability of the uuddsQ system within
the chiral constituent quark model \cite{GPP96} used previously in the study
of the H-particle \cite{SPG97} and of heavy tetraquarks as well \cite{PSGR97}.
 An essential difference with respect to Refs 
\cite{LSB93,ZSGR86,SBS93,BS97} is that the spin-spin interaction of 
\cite{GPP96} is flavour-dependent. In Ref.\cite{PSGR97}
we found that the Goldstone boson
exchange interaction between quarks binds strongly both the $cc\bar{q}\bar{q}$
system and the $bb\bar{q}\bar{q}$ system. Within conventional models based 
on one-gluon exchange, the  $cc\bar{q}\bar{q}$ was found unstable and
 $bb\bar{q}\bar{q}$ stable \cite{SBS93}.

In section 2, we establish the basis states required by the internal 
symmetries of the system under discussion and, based on simple group 
theory arguments, we indicate the most important basis states for a
given isospin I and total angular momentum J. In section 3, we briefly 
describe the Hamiltonian of the chiral constituent quark model used in the
calculations. In section 4, we present our results and the last section is
devoted to a summary.

\section{BASIS STATES}
In principle the GBE interaction contains all pairs ij of
particles. But the exchange of a heavy pseudoscalar meson, between a light
and a heavy quark Q, can in practice be neglected \cite{PSGR97}. For
example, it was explicitly shown in Ref.\cite{GR96+} that the dominant 
contribution
to the masses of C=+1 charmed baryons is due to meson exchange
between light quarks and that the exchange of $D$ and $D_s$ mesons is
negligible, of the order of few MeV. Inasmuch as the quark-quark interaction
\cite{GR96,GPP96} is inverse proportional to the masses of the interacting
quarks, the exchange of $B$ or $B_s$ meson can further be neglected. 
Neglecting the contribution of $D, D_s, B$
and $B_s$ mesons,
the GBE interaction in the uuddsQ system, with Q=c or b, reduces to
the interaction between light quarks q=u, d or s.
 
We adopt this point of view, i.e. we neglect the contribution of heavy 
meson-exchange. Then similarly to the study of the H-particle or of the 
NN system \cite{NN97} it
is useful to have a qualitative insight about the uuddsQ system by first
considering a schematic quark-quark interaction which simplifies the GBE
interaction of Ref.\cite{GPP96}, by removing its radial dependence. The
schematic interaction reads:
\begin{equation}
 V_{\chi} = - C_{\chi} \sum_{i<j}  \lambda_{i}^{F} . \lambda_{j}^{F}
\vec{\sigma}_i . \vec{\sigma}_j ,
\label{opFS}
\end{equation}
where $\lambda_{i}^{F}$ (F=1,2,...,8) are the quark-flavour Gell-Mann
matrices (with an implied summation over F) and $\vec{\sigma}$ are the spin
matrices. The minus sign of the interaction (\ref{opFS}) is related to the
sign of the short-range part of the GBE interaction, crucial for the
hyperfine splitting in baryon spectroscopy. This feature of the short-range
part of the GBE interaction is clearly discussed at length by Glozman and
Riska \cite{GR96}. A typical order of magnitude
for the constant $C_{\chi}$ is about 30 MeV.

In order to calculate the
expectation value of (\ref{opFS}) for the remaining light pentaquark system
$q^5$ we have to give a classification of its states. In the colour space this
system is described by the state $[221]_C$ compatible with the colourless
state $[222]_C$ of $q^5Q$. Here and below [f] stands for the corresponding
Young diagram in the colour (C), spin (S) or flavour (F) space. 

We first assume that u,d and s are identical. If the quarks are all in the
ground state the orbital part of the wave function is symmetric. Then the
only $q^5$ flavour-spin state allowed by the Pauli principle is
$[32]_{FS}$. By using inner product rules \cite{book} one can find the
flavour $[f]_F$ and spin $[f]_S$ symmetries compatible with $[32]_{FS}$. These
are listed in Table 1, together with the corresponding isospin I and spin S
associated with these states. We also give the total angular momentum
$\vec{J} = \vec{S} + \vec{S}_Q$ of the $q^5Q$ system. The last column
reproduces the expectation value of (\ref{opFS}) in units of $C_{\chi}$. This
has been calculated using the formula given in the Appendix A of
Ref.\cite{NN97}, containing the Casimir operators of $SU(6)_{FS}$,
$SU(3)_{F}$ and $SU(2)_S$. The multiplicity of a given IJ state is consistent
with Table 3 (Y=2/3) of Ref.\cite{LSB93}. From Table 1, one can see that the
most favourable candidate for stability (the most negative eigenvalue of
(\ref{opFS})) should have I=0 (flavour symmetry $[221]_F$) and J=0 or 1. In
the numerical calculations given below, different flavour symmetries will be
mixed by the GBE interaction (\ref{opFS}).

 In the diagonalization procedure of the GBE Hamiltonian given below, 
we truncate
the basis for various sectors IJ, by retaining only the lowest states. The
mixture with the others is expected to be small. Note also that the flavour
spin interaction of type (\ref{opFS}) does not mix
$[f]_S \neq [f']_S$. In numerical calculations, we therefore restrict the
basis states to the following ones :

\underline{I=0, J=0 or 1}
\begin{equation}
|1> = |[221]_F [32]_S>;  |2> = |[32]_F [32]_S>
\label{bas1}
\end{equation}

\underline{I=1, J=0 or 1}
\begin{equation}
|1> = |[311]_F [32]_S>;  |2> = |[32]_F [32]_S>
\label{bas2}
\end{equation}

\underline{I=2, J=0 or 1}
\begin{equation}
|1> = |[41]_F [32]_S>;  |2> = |[5]_F [32]_S>
\label{bas3}
\end{equation}

\section{HAMILTONIAN}
The Hamiltonian to be diagonalized is \cite{GR96+} :
\begin{equation}
H= \sum_i m_i + \sum_i \frac{\vec{p}_{i}^{2}}{2m_i} - \frac {(\sum_i
\vec{p}_{i})^2}{2\sum_i m_i} + \sum_{i<j} V_{conf}(r_{ij}) + \sum_{i<j}
V_\chi(r_{ij})
\label{ham}
\end{equation}
with the linear confining interaction :
\begin{equation}
 V_{conf}(r_{ij}) = -\frac{3}{8}\lambda_{i}^{c}\cdot\lambda_{j}^{c} \, C
\, r_{ij}
\label{conf}
\end{equation}
and the spin-spin component of the GBE interaction in its $SU_F(3)$ form :
\begin{eqnarray}
V_\chi(\vec r_{ij})
&=&
\left\{\sum_{F=1}^3 V_{\pi}(\vec r_{ij}) \lambda_i^F \lambda_j^F \right.
\nonumber \\
&+& \left. \sum_{F=4}^7 V_{\rm K}(\vec r_{ij}) \lambda_i^F \lambda_j^F
+V_{\eta}(\vec r_{ij}) \lambda_i^8 \lambda_j^8
+V_{\eta^{\prime}}(\vec r_{ij}) \lambda_i^0 \lambda_j^0\right\}
\vec\sigma_i\cdot\vec\sigma_j,
\label{VCHI}
\end{eqnarray}
 
\noindent
with $\lambda^0 = \sqrt{2/3}~{\bf 1}$, where $\bf 1$ is the $3\times3$ unit
matrix. The interaction (\ref{VCHI}) contains $\gamma = \pi, K, \eta$ and
$\eta'$ exchanges and the form of $V_{\gamma}(r_{ij})$ is given explicitly in
Ref.\cite{GPP96} as the sum of two distinct contributions : a Yukawa type
potential containing the mass of the exchanged meson and a short-range
contribution, of opposite sign, the role of which is crucial in baryon
spectroscopy. For a given meson $\gamma$, the meson exchange potential is :
\begin{equation}V_\gamma (\vec r_{ij})=
\frac{g_\gamma^2}{4\pi}\frac{1}{3}\frac{1}{4m_i m_j}
\{\mu_\gamma^2\frac{e^{-\mu_\gamma r_{ij}}}{ r_{ij}}- \frac {4}{\sqrt {\pi}}
\alpha^3 \exp(-\alpha^2(r-r_0)^2)\}
,  \hspace{5mm} (\gamma = \pi, K, \eta, \eta' )
\label{POINT} \end{equation}

For the Hamiltonian (\ref{ham})-(\ref{POINT}), we use the
parameters of Ref.\cite{GPP96}. These are :
$$\frac{g_{\pi q}^2}{4\pi} = \frac{g_{\eta q}^2}{4\pi} = 
\frac{g_{Kq}^2}{4\pi}= 0.67;\,\,
\frac{g_{\eta ' q}^2}{4\pi} = 1.206$$
$$r_0 = 0.43 \, {\rm fm}, ~\alpha = 2.91 \, {\rm fm}^{-1},~~
 C= 0.474 \, {\rm fm}^{-2}, m_{u,d} = 340 \, {\rm MeV}. $$
\begin{equation}
 \mu_{\pi} = 139 \, {\rm MeV},~ \mu_{\eta} = 547 \, {\rm MeV},~
\mu_{\eta'} = 958 \, {\rm MeV},~ \mu_{K} = 495 \, {\rm MeV}.
\label{PAR} \end{equation}
They provide a very satisfactory description of low-lying nonstrange baryons,
extended to strange baryons in \cite{semrel} in a fully dynamical
three-body calculations as well. The latter reference gives $m_s=0.440$ GeV.
For the masses of the heavy quarks we take $m_c=1.35$ GeV, $m_b=4.66$ GeV in
agreement with Ref.\cite{PSGR97} where these masses are adjusted to reproduce
the average mass $\bar{M}=(M + 3 M^*)/4$ of M = D and B mesons respectively.
Note that within the spirit of the model of Glozman and Riska there is no meson
exchange between a quark and a antiquark.

\section{RESULTS}
First we discuss the spin S=1/2 baryons needed to calculate the threshold
energy. Using the Hamiltonian (\ref{ham})-(\ref{POINT}), we have performed 
variational estimates with a general wave function of the 
form $\psi \sim \exp{[-(ax^2+by^2)]}$
where  $\vec{x} = \vec{r}_1 - \vec{r}_2$, $\vec{y} = (\vec{r}_1 + \vec{r}_2 -
2 \vec{r}_3)/\sqrt{3}$. To
the nucleon mass, only $\pi, \eta$ and $\eta'$ exchange contribute. To the 
mass of $\Lambda$ or $\Sigma$, there is a contribution from K-exchange as well.
In the case of heavy baryons $\Lambda_c,\Lambda_b,\Sigma_c, \Sigma_b, \Xi_c$
and $\Xi_b$ we neglect any meson exchange between a light and a heavy quark,
in the spirit of the above discussion.
 The results are presented in Table 2, where for N,
$\Lambda$ and $\Sigma$ we made the simplification a=b. The expectation values
lie typically about 50 MeV above the experimental value. The results for
$\Sigma$ could be improved by taking a$\neq$b and those for $\Lambda_b$ made
more realistic by tuning the mass of $m_b$. But, for the present purpose, as
will be seen below, these estimates are quite satisfactory.

 For the hexaquarks discussed here, it is useful to introduce the following
system of Jacobi coordinates (where i=1,2,...,5 are associated with light
quarks and i=6 with the heavy one):
\begin{equation}
\begin{array}{ccl}
\vec{x} &=& \vec{r}_1 - \vec{r}_2 \\
\vec{y} &=& \vec{r}_3 - \vec{r}_4 \\
\vec{z} &=& \frac{1}{\sqrt{2}} (\vec{r}_1 + \vec{r}_2 - \vec{r}_3 -
\vec{r}_4)\\
\vec{t} &=& \frac{1}{\sqrt{10}} (\vec{r}_1 + \vec{r}_2 + \vec{r}_3 +
\vec{r}_4 - 4 \vec{r}_5)\\
\vec{w} &=& \frac{1}{\sqrt{15}} (\vec{r}_1 + \vec{r}_2 + \vec{r}_3 +
\vec{r}_4 + \vec{r}_5 - 5 \vec{r}_6)\\
\vec{R}_{CM} &=& (m\vec{r}_1 + m\vec{r}_2 + m\vec{r}_3 +
m\vec{r}_4 + m_s \vec{r}_5 + m_Q \vec{r}_6)/(4m+m_s+m_Q) 
\end{array}
\end{equation}
Moreover, in the kinetic term only, we use the average mass :
\begin{equation} 
\bar{m} = (4m+m_s)/5
\label{moyen}
\end{equation}
for all light quarks.

By assuming a ground state variational wave function of the form:
\begin{equation}
\psi = (\frac{a}{\pi})^3 (\frac{b}{\pi})^{3/4}
\exp{[-\frac{a}{2}(x^2+y^2+z^2+t^2)-\frac{b}{2}w^2]}
\label{wf}
\end{equation}
the expectation value $E_0$ of the spin-independent part of the Hamiltonian
becomes:
\begin{equation}
\begin{array}{ccc}
E_0 & = & \displaystyle \frac{6}{\bar{m}} \frac{\hbar^2c^2}{2a^2} +
(\frac{1}{\bar{m}} +
\frac{5}{m_b}) \frac{\hbar^2c^2}{8b^2} \\
 &  & + \displaystyle \frac{4}{5} C [10 \sqrt{\frac{1}{\pi a}} + 5
\sqrt{\frac{1}{5\pi} (\frac{2}{a} + \frac{3}{b})} ] 
\end{array}
\end{equation}
For the matrix elements of the spin-dependent part, the fractional parentage
technique \cite{book} has been used. Details are given in Appendix A.

In Table 3, we present results from the diagonalization of the Hamiltonian
(\ref{ham})-(\ref{POINT}) for the bases (\ref{bas1})-(\ref{bas3}). The 
column-heading M represents the lowest expectation value obtained 
in that sector.
The next column gives the lowest theoretical threshold compatible with a 
given IJ. We also indicate the next to the lowest threshold whenever it is 
close to the lowest one. Note that using the experimental masses, Table 2, 
last column, the two thresholds may interchange their position. The
last column is the lowest eigenvalue from which the threshold mass $M_T$
obtained from Table 2 has been subtracted. The lowest eigenvalue is the
equilibrium value obtained by minimizing with respect to the variational
parameters a and b of (\ref{wf}). In each case it turns out that b at
equilibrium is approximately equal to the value of b given in Table 2,
associated to the heavier of the threshold baryons. At equilibrium, we
 also find that the off-diagonal matrix elements of the GBE interaction are
typically one order of magnitude smaller than the diagonal ones so that the
lowest state does not change much through the coupling to the next state. 
A typical change is of a few MeV. This
also proves that the truncation of the bases as in (\ref{bas1})-(\ref{bas2})
is safe. The smallest  $M-M_T$ corresponds to IJ=00 or 01, as expected from
the discussion following Table 1. In all cases
$M-M_T$ is positive and very large which means that none of the considered
system is stable against strong decays. Actually, there is a substantial
amount of repulsion, similar to the case of the H-particle \cite{SPG97}. Thus
within the GBE model used here, the heavy compact hexaquark uuddsb is highly
unstable, contrary to the findings of Ref.\cite{LRP96} or \cite{LSB93}. In the
latter reference the system with I=0, J=2 is bound by 13.8 MeV for the most
favourable choice of the model parameters.

The amount of repulsion found depends, of course, on the approximations used,
and in particular on the treatment of the kinetic energy. A better
treatment, where instead of the mass average (\ref{moyen}) the kinetic energy
 is expressed in terms of the average of the inverse of the reduced masses
, may slightly decrease the kinetic contribution but certainly will
not change the above conclusion. The expectation is that the incorporation of 
the D or B meson exchange will not change the conclusion either.
But in cases where uuddsQ is found to have a mass close to the lowest 
threshold, as for example Ref\cite{LSB93}, a proper treatment of the kinetic 
energy is much more important in drawing a conclusion about stability under 
strong interactions. In a shell model description where one assumes that all
the quarks are in an s state, as in \cite{LSB93}, one can make an estimate of
the kinetic energy of a system of six identical quarks from which one can
subtract the kinetic energy of two separate clusters of three quarks each.
This gives $3/4 \hbar \omega$ \cite{SPG97}, i.e. a positive contribution, 
which may counterbalance the binding found in Ref.\cite{LSB93}. 

One can also raise the question whether or not by increasing the number of
heavy quarks the stability would increase. In \cite{PSGR97}, we
also investigated the system qqqqQQ in the Glozman et al. model \cite{GPP96},
by using a similar procedure. There the most favourable configuration has I=0,
J=1. For Q=c we obtained $M-M_T = 0.523$ GeV and for Q=b, $M-M_T = 0.515$ GeV.
In both cases $M_T$ corresponds to the lowest threshold qQQ + qqq where 
m(ccu) = 3.514 GeV and m(bbu) = 10.066 GeV. Therefore, these systems are 
also unbound in a compact configuration. 

\section{SUMMARY}
 In a chiral constituent quark model which successfully describes the
light, strange and the presently known charmed and b-baryons, we have
calculated the mass $M$ of the hexaquarks $H_c$ (uuddsc) and $H_b$ (uuddsb) for
various IJ sectors and compared it to the mass $M_T$ of the corresponding
lowest threshold. We found that the smallest $M-M_T$ value is associated
to the IJ=00 or 01 sector. The quantity $M-M_T$ is always positive and of 
the order of few hundreds MeV. This indicates that $H_c$ and $H_b$ cannot 
exist as compact systems.
 
However,
the existence of a weakly bound, molecular-type heavy hexaquark system, like
the deuteron, cannot be excluded. The GBE interaction \cite{GR96} generates a
long-range attraction due to its Yukawa-potential tail and in principle, it can also
produce a medium-range attraction from correlated
two-pseudoscalar meson exchange. It is certainly interesting to pursue
investigations in this direction, in a dynamical approach as the resonating
group or the generator coordinate method, by incorporating six-quark states
with orbital excitations, as in the NN case \cite{NN97}.

\acknowledgments
We are grateful to M. Genovese and J.-M. Richard for many valuable discussions.

\appendix
\section{}
The calculation of the matrix elements of the interaction potential
(\ref{VCHI}) between five light quarks is based on the fractional parentage
technique described in Ref.\cite{book}. Through this technique each five-body
matrix element reduces to a linear combination of two-body matrix elements of
the pair, 4 and 5, of quarks. This is possible due to the fact that each part
(orbital, spin, flavour, colour) of the wave function is written as a sum of 
products of the first three and of the pair (45) wave functions. These 
functions have a definite permutation symmetry [f] and [f'] respectively, 
where [f']=[2] or [11]. The coefficients of these linear combinations have 
been obtained from the isoscalar factors of the Clebsch-Gordan coefficients
of $S_5$ as calculated in Ref.\cite{PS96}. 

The calculation of the spin-spin matrix elements is trivial. Below we give
the flavour wave functions which we derived in the
Rutherford-Young-Yamanouchi representation, where the pair 45 is either in a
symmetric or an antisymmetric state, as mentioned above. 
We denote by $p$ and $q$ the row of the
5th and 4th particle in a Young tableau and by $\bar{pq}$ and $\tilde{pq}$ a
symmetric and an antisymmetric state respectively. Then the uudds states
required in these calculations are:
\begin{equation}
\begin{array}{ccc}
|[5]\bar{11}> & = & \sqrt{\frac{1}{5}}\psi_{[3]}(uud)\phi_{[2]}(ds) +
\sqrt{\frac{1}{5}}\psi_{[3]}(udd)\phi_{[2]}(us) \\
 & & + \sqrt{\frac{2}{5}}\psi_{[3]}(uds)\phi_{[2]}(ud) +
\sqrt{\frac{1}{10}}\psi_{[3]}(dds)\phi_{[2]}(uu) \\
 & & + \sqrt{\frac{1}{10}}\psi_{[3]}(uus)\phi_{[2]}(dd)
\end{array}
\end{equation}

\begin{equation}
\begin{array}{ccc}
|[41]\bar{11}> & = &
-\sqrt{\frac{4}{6}}\psi_{[21]}^{\rho,\Sigma_0}(uds)\phi_{[2]}(ud) +
\sqrt{\frac{1}{6}}\psi_{[21]}^{\rho}(dsd)\phi_{[2]}(uu) \\
 & & + \sqrt{\frac{1}{6}}\psi_{[21]}^{\rho}(usu)\phi_{[2]}(uu)
\end{array}
\end{equation}

\begin{equation}
\begin{array}{ccc}
|[41]\bar{12}> & = & \sqrt{\frac{3}{10}}\psi_{[3]}(uud)\phi_{[2]}(ds) +
\sqrt{\frac{3}{10}}\psi_{[3]}(udd)\phi_{[2]}(us) \\
 & & - \sqrt{\frac{4}{15}}\psi_{[3]}(uds)\phi_{[2]}(ud) -
\sqrt{\frac{1}{15}}\psi_{[3]}(dds)\phi_{[2]}(uu) \\
 & & -\sqrt{\frac{1}{15}}\psi_{[3]}(uus)\phi_{[2]}(dd)
\end{array}
\end{equation}

\begin{equation}
|[41]\tilde{12}>  =  -\sqrt{\frac{1}{2}}\psi_{[3]}(uud)\phi_{[11]}(ds) -
\sqrt{\frac{1}{2}}\psi_{[3]}(udd)\phi_{[11]}(us)
\end{equation}

\begin{equation}
\begin{array}{ccc}
|[32]\bar{22}> & = & \frac{1}{3}\psi_{[3]}(uud)\phi_{[2]}(ds) 
-\frac{1}{3}\psi_{[3]}(udd)\phi_{[2]}(us) \\
 & & - \frac{\sqrt{2}}{3}\psi_{[3]}(uds)\phi_{[2]}(ud) +
\frac{1}{\sqrt{2}}\psi_{[3]}(dds)\phi_{[2]}(uu) \\
 & & +\frac{1}{3\sqrt{2}}\psi_{[3]}(uus)\phi_{[2]}(dd)
\end{array}
\end{equation}

\begin{equation}
\begin{array}{ccc}
|[32]\bar{12}> & = & \frac{2}{3} \psi_{[21]}^{\rho}(udu)\phi_{[2]}(ds) +
\frac{1}{3} \psi_{[21]}^{\rho}(dud)\phi_{[2]}(us) \\
 & & - \frac{\sqrt{2}}{3} \psi_{[21]}^{\rho}(uds)\phi_{[2]}(ud) +
\frac{\sqrt{2}}{3} \psi_{[21]}^{\rho}(usu)\phi_{[2]}(dd)
\end{array}
\end{equation}

\begin{equation}
|[32]\tilde{12}> = \sqrt{\frac{1}{3}} \psi_{[21]}^{\rho}(dud)\phi_{[11]}(us)
- \sqrt{\frac{2}{3}} \psi_{[21]}^{\rho}(uds)\phi_{[11]}(ud)
\end{equation} 

\begin{equation}
\begin{array}{ccc}
|[311]\bar{11}> &=& \psi_{[111]}(uds)\phi_{[2]}(ud)
\end{array}
\end{equation}

\begin{equation}
\begin{array}{ccc}
|[311]\bar{13}> & = & \frac{\sqrt{15}}{10}
\psi_{[21]}^{\rho}(udu)\phi_{[2]}(ds) -
\frac{\sqrt{15}}{10} \psi_{[21]}^{\rho}(dud)\phi_{[2]}(us) \\
 & & - \sqrt{\frac{1}{10}} \psi_{[21]}^{\rho,\Lambda^0}(uds)\phi_{[2]}(ud) +
\sqrt{\frac{3}{10}} \psi_{[21]}^{\rho}(dsd)\phi_{[2]}(uu) \\
 & & -\sqrt{\frac{3}{10}} \psi_{[21]}^{\rho}(usu)\phi_{[2]}(dd)
\end{array}
\end{equation}

\begin{equation}
\begin{array}{ccc}
|[311]\tilde{13}> & = & -\frac{1}{2} \psi_{[21]}^{\rho}(udu)\phi_{[11]}(ds) +
\frac{1}{2} \psi_{[21]}^{\rho}(dud)\phi_{[11]}(us) \\
 & & - \sqrt{\frac{1}{2}} \psi_{[21]}^{\rho,\Sigma_0}(uds)\phi_{[11]}(ud) 
\end{array}
\end{equation}

\begin{equation}
\begin{array}{ccc}
|[311]\tilde{23}> & = & \sqrt{\frac{2}{5}}
\psi_{[3]}(uud)\phi_{[11]}(ds) -
\sqrt{\frac{2}{5}} \psi_{[3]}(udd)\phi_{[11]}(us) \\
 & & + \sqrt{\frac{1}{5}} \psi_{[3]}(uds)\phi_{[11]}(ud) 
\end{array}
\end{equation}

\begin{equation}
\begin{array}{ccc}
|[221]\bar{23}> & = & \frac{\sqrt{2}}{4}
\psi_{[21]}^{\rho}(udu)\phi_{[2]}(ds) +
\frac{\sqrt{2}}{4} \psi_{[21]}^{\rho}(dud)\phi_{[2]}(us) \\
 & & - \frac{1}{2} \psi_{[21]}^{\rho,\Sigma^0}(uds)\phi_{[2]}(ud) -
\frac{1}{2} \psi_{[21]}^{\rho}(dsd)\phi_{[2]}(uu) \\
 & & - \frac{1}{2} \psi_{[21]}^{\rho}(usu)\phi_{[2]}(dd)
\end{array}
\end{equation}

\begin{equation}
\begin{array}{ccc}
|[221]\tilde{23}> & = & -\frac{\sqrt{6}}{4}
\psi_{[21]}^{\rho}(udu)\phi_{[11]}(ds) -
\frac{\sqrt{6}}{4} \psi_{[21]}^{\rho}(dud)\phi_{[11]}(us) \\
 & & - \frac{1}{2} \psi_{[21]}^{\rho,\Lambda^0}(uds)\phi_{[11]}(ud) 
\end{array}
\end{equation}

\begin{equation}
\begin{array}{ccc}
|[221]\tilde{12}> &=& \psi_{[111]}(uds)\phi_{[11]}(ud)
\end{array}
\end{equation}

The states $\psi_{[2]}(ab)$ and $\psi_{[11]}(ab)$ are the symmetric and 
antisymmetric two-particle states. The  $\psi_{[3]}(abc)$ is the symmetric
three particle states. For mixed symmetry states  $\psi_{[21]}^{\rho}$ some
care should be taken. As usually \cite{book} one has :
\begin{equation}
 \psi_{[21]}^{\rho}(udu) = \frac{1}{2} (udu-duu)
\label{rho}
\end{equation}
\begin{equation}
 \psi_{[21]}^{\rho,\Lambda_0}(uds) = \frac{1}{\sqrt{12}}(2 uds - 2 dus + sdu
-sud + usd  - dsu)
\label{lam}
\end{equation}
\begin{equation}
 \psi_{[21]}^{\rho,\Sigma_0}(uds) = -\frac{1}{2}(usd + dsu -sdu -sud)
\label{sig}
\end{equation}

However, it turns out that the states $|[32]\bar{12}>$ and $|[32]\tilde{12}>$
contain the function $\psi^{\rho}_{[21]}$, the definition of which is :
\begin{equation}
\psi^{\rho}_{[21]}(uds) = \frac{\sqrt{3}}{2} \psi^{\rho,\Lambda_0}_{[21]} -
\frac{1}{2} \psi^{\rho,\Sigma_0}_{[21]}
\label{tot}
\end{equation}
i.e. a linear combination of (\ref{lam}) and (\ref{sig}). While the states
(\ref{rho})- (\ref{sig}) have a definite isospin the state (\ref{tot}) is a
 mixture of I=0 and I=1. Thus the states $|[32]\bar{12}>$ and 
$|[32]\tilde{12}>$ do not have a definite isospin. Therefore one has to project
into a specific value of I in the calculation of matrix element of these two
states. Calculation with or without projection indicate a difference of few 
MeV which is insignificant in the context of the present study.

\begin{table}
\caption[Expectation values]{\label{expectation} Expectation value of the 
operator (\ref{opFS}) in units of $C_\chi$ for all flavour $[f]_F$ and 
spin $[f]_S$ symmetries compatibles with $[32]_{FS}$. The corresponding 
spin S and isospin I, together with the total angular momentum J are also 
given.}
\begin{tabular}{dddddd}
 $[f]_F$ & $[f]_S$  & S & J &
 I  & $<V_{\chi}>$ \\
\tableline
$[221]$ & [41] & 3/2 & 1,2 & 0 & -12 \\ 
$[221]$ & [32] & 1/2 & 0,1 & 0 & -16 \\
$[311]$ & [32] & 1/2 & 0,1 & 1 & -12 \\
$[32]$ & [32] & 1/2 & 0,1 & 0,1 & -8 \\
$[311]$ & [41] & 3/2 & 1,2 & 1 & -8 \\
$[32]$ & [41] & 3/2 & 1,2 & 0,1 & -4 \\
$[41]$ & [32] & 1/2 & 0,1 & 1,2 & -2 \\
$[41]$ & [41] & 3/2 & 1,2 & 1,2 & 2 \\
$[32]$ & [5] & 5/2 & 2,3 & 0,1 & 8/3 \\
$[5]$ & [32] & 1/2 & 0,1 & 2 & 8 \\
\end{tabular}
\end{table}

\begin{table}
\renewcommand{\arraystretch}{1.5}
\parbox{18cm}{\caption[Resultsbaryons]{\label{baryons} Variational solution
of the Hamiltonian (\ref{ham})-(\ref{VCHI}) for spin S=1/2 low lying baryons 
compared to experimental masses}}
\begin{tabular}{c@{\quad}|c@{\quad}|c@{ }|c@{ }|c}
Baryon & \multicolumn{2}{c@{ }|}{Variational parameters (fm)} & 
Expectation value
& Experimental mass (GeV) \\
   & a & b & (GeV) & \hspace{4mm} \cite{PDG96} \\
\tableline
N & 0.4376 & & 0.9696 & 0.940 \\
$\Lambda$ & 0.4486 &  & 1.1654 & 1.1156 \\
$\Sigma$ & 0.4625 &  & 1.2354 & 1.193 \\
$\Lambda_c$ & 0.4683 & 0.7099 & 2.3268 & 2.2849 \\
$\Sigma_c$ & 0.6320 & 0.7201 & 2.4889 & 2.452 \\
$\Xi_c$    & 0.5705 & 0.6967 & 2.5494 & 2.470 \\
$\Lambda_b$ & 0.4678 & 0.6509 & 5.6147 & 5.641 \\
$\Sigma_b$ & 0.6292 & 0.6623 & 5.7775 & ? \\
$\Xi_b$ &    0.5683 & 0.6339 & 5.83629 & ? \\
\end{tabular}
\end{table}

\begin{table}
\renewcommand{\arraystretch}{1.5}
\parbox{18cm}{\caption[Results]{\label{results} Masses M and mass
differences $M-M_T$ ($M_T$ - the threshold mass) of heavy hexaquark systems
uuddsQ (Q=c or b) of given isospin I and total angular momentum J}}
\begin{tabular}{cccccc}
System & I & J  & M & Threshold & $M - M_T$ \\
       &   &   & (GeV) &       & (GeV)  \\
\tableline
uuddsc & 0 & 0,1 & 4.144 &  $\begin{array}{ccc} N & + & \Xi_{c} \\
\Lambda &+& \Lambda_c \end{array}$  & $\begin{array}{c} 0.625 \\ 0.652 
\end{array}$  \\
\tableline
   & 1 & 0,1 & 4.304 &  $\begin{array}{ccc} \Sigma &+& \Lambda_c  \\ 
N & + & \Xi_c \end{array}$ 
 &  $\begin{array}{c} 0.742 \\ 0.785
 \end{array}$  \\
\tableline
   & 2 & 0,1 & 4.496 & $\Sigma + \Sigma_c$ & 0.772 \\
       &   &   &  &       &   \\
\tableline
\tableline
uuddsb & 0 & 0,1 & 7.425 &  $\begin{array}{ccc} N & + & \Xi_b \\
\Lambda &+& \Lambda_b \end{array}$  &  $\begin{array}{c} 0.619 \\ 0.645 
\end{array}$  \\
\tableline
   & 1 & 0,1 & 7.586 & $\begin{array}{ccc} \Sigma &+& \Lambda_b  \\ 
N & + & \Xi_b \end{array}$
&  $\begin{array}{c} 0.736 \\ 0.780
 \end{array}$  \\
\tableline
   & 2 & 0,1 & 7.780 & $\Sigma + \Sigma_b$ & 0.767 \\
\end{tabular}
\end{table}
\end{document}